\documentclass[conference]{IEEEtran}
\ifCLASSINFOpdf

\else

\fi

\usepackage{graphicx}
\usepackage{amsthm}
\usepackage{subfigure}
\usepackage{amssymb, amsmath, amsfonts, epsfig, latexsym, times}
\usepackage{bm}
\usepackage{cite}
\usepackage{flushend}
\usepackage{booktabs}
\usepackage{multirow}
\usepackage{subeqnarray}
\usepackage{cases}
\usepackage{mathrsfs}
\newcommand{\tabincell}[2]{\begin{tabular}{@{}#1@{}}#2\end{tabular}}
\usepackage{algorithmic}
\usepackage{algorithm}
\begin{document}

\title{Joint User Association and Downlink Beamforming for Green Cloud-RANs with Limited Fronthaul}
\author{\IEEEauthorblockN{Zhi Yu, Ke Wang, Hong Ji, Xi Li, Heli Zhang}
\IEEEauthorblockA{Key Laboratory of Universal Wireless Communications, Ministry of Education\\Beijing University of Posts and Telecommunications\\
Beijing, P.R. China\\
Email: zhiyu, wangke, jihong, lixi, zhangheli$\{$@bupt.edu.cn$\}$}

}

\maketitle
\bibliographystyle{plain}

\begin{abstract}
\boldmath
With the explosive growth of smart devices and mobile data traffic, limited fronthaul capacity has become a notable bottleneck of green communication access networks, such as cloud radio access networks(C-RANs). In this paper, we proposed a joint user association and downlink beamforming scheme for green C-RANs to minimize the network power consumption with the limited fronthaul links. We first formulate the design problem as a mixed integer nonlinear programming (MINLP), and then transformed the MINLP problem into a mixed integer second-order cone program (MI-SOCP) which is a convex program when the integer variables are fixed. By relaxing the integer variables to continuous ones, an inflation algorithm, which can be finished within polynomial time, was proposed to solved the problem. The simulation results are presented to validate the effectiveness of our proposed algorithm compared with the the scheme adopted by LTE-A.
\end{abstract}


\IEEEpeerreviewmaketitle

\section{Introduction}
Cloud radio access network (C-RAN) has recently been proposed as a promising network architecture to jointly improve energy efficiency and spectrum efficiency, and reduce both the capital expenses and operation expenses \cite{Introd4}. In C-RANs, baseband data and channel state information (CSI) are processed in a central processor called Baseband Unit (BBU) pool and shared among the densely deployed remote radio heads (RRHs) via fronthaul links, which allows the RRHs to cooperatively transmit the data to the mobile users (MUs) \cite{Introd3}. Unfortunately, with the dense deployment of RRHs, the overall network power consumption, including both transmit and circuit power consumption \cite{power3}, will increase significantly. Therefore, green C-RAN has quickly attracted wide attention \cite{Introd11,nof1,nof2}.

The authors in \cite{Introd11} address the problem of downlink beamforming to improve energy efficiency of C-RANs by using weighted mixed norm minimization. A three-stage algorithm based on the group-sparsity inducing norm is presented to minimize the power consumption for multicast C-RANs \cite{nof1}. Combining virtualized network resources and virtualized functional entities of baseband processing, the authors in \cite{nof2} propose an energy-saving scheme for C-RANs based on formation of Virtual Base Station. However, these excellent works do not take into account limitation of fronthaul links, which is becoming the a bottleneck for realizing the potential performance gain of C-RANs \cite{front1}. Moreover, with the huge number of MUs involved in the network, enormous baseband signals and signaling overheads are required to be transmitted in fronthaul links, which incurs that the fronthaul limited problem becomes more severe. Hence, finding a solution of network power minimization problem under limited fronthaul is critical to achieve green communication and commercial deployment of C-RANs.

Some other excellent works have been done on energy saving with constrained fronthaul links in CRANs. The authors in \cite{Introd13} investigate the tradeoff between total transmit power and sum backhaul capacity over all BSs in a network MIMO system. To minimize downlink transmit power, a efficient algorithm with constraints on fronthaul capacity is presented in \cite{Introd14}. For transmit power minimization, the authors in \cite{Introd15} jointly consider coordinated beamforming and admission control design in fronthaul constrained C-RANs. Nevertheless, all these works assume that all the RRHs are involved to cooperatively transmit data to MUs, i.e., the RRH sleeping mechanism which is closely related to the user association status is not considered. Therefore, the schemes proposed by these works can not efficiently reduce the circuit power consumption, which accounts for a large part of network power consumption \cite{circuit}.

In this paper, we investigate the network power minimization problem in C-RAN with limited fronthaul. This design problem jointly considers user association and downlink beamforming, which dominate circuit power consumption and transmit power consumption, respectively.
to be more specific, we first characterize the problem as a mixed integer nonlinear program (MINLP), which is hard to be solved because of the features of the user association. Then, we transform the problem into a mixed integer second-order cone program(MI-SOCP) by making the objective function plus a sufficient small variable. Finally, we relax the integer variables to continuous ones and adopt a low complexity inflation algorithm to obtain the suboptimal solution. Numerical results verify the effectiveness of the proposed scheme by compared with the scheme adopted by LTE-A.

The rest of this paper is organized as follows. Section \uppercase\expandafter{\romannumeral2} presents the system and power model. In Section \uppercase\expandafter{\romannumeral3}, the network power consumption minimization problem is formulated as a MI-SOCP, followed by some analisis. Section \uppercase\expandafter{\romannumeral4} presents the low complexity inflation algorithm to obtain the suboptimal solution. Simulation results and discussions are given in section \uppercase\expandafter{\romannumeral5}. Finally, we conclude this paper in section \uppercase\expandafter{\romannumeral6}.
\section{System and Power Model}
\subsection{System Model}
\begin{figure}[!t]
\centering
\includegraphics[width=3.05in]{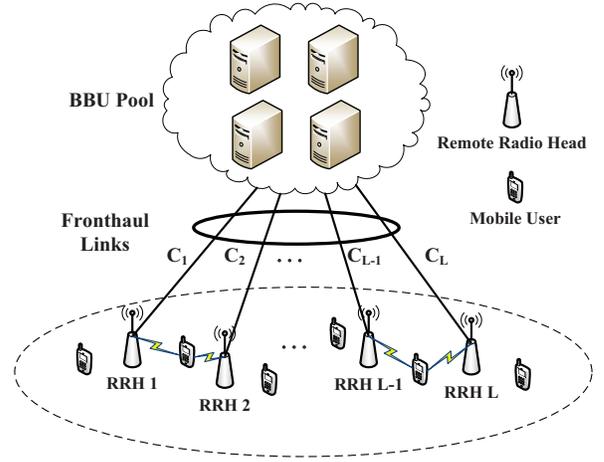}
\caption{An example of downlink C-RAN with limited-capacity fronthaul links.}\label{tab:CRAN}
\end{figure}

We consider a downlink Cloud-RAN with $L$ multiple-antenna RRHs and $K$ single-antenna MUs, where the $l$-th RRH with $N_{l}$ antennas is connected to a BBU Pool by a limited fronthaul link, as shown in Fig.\ref{tab:CRAN}. It is assumed that all user data and the perfect CSI is available at the BBU pool. We define $\mathcal{L}=\{1,2,\ldots,L\}$ and $\mathcal{K}=\{1,2,\ldots,K\}$ as the set of RRH and MU indices, respectively. In a beamformer design problem, the baseband signals of RRH $l$ can be expressed as
\begin{equation}
\begin{aligned}
x_{l}=\sum\limits_{k=1}^{K}\mathbf{w}_{l,k}s_{k}, \forall l\in\mathcal{L}
\end{aligned}
\end{equation}
where $\mathbf{w}_{l,k}\in\mathbb{C}^{N_{l}\times1}, \forall l\in \mathcal{L}, \forall k\in \mathcal{K}$ is the beamforming vector at RRH $l$ for MU $k$ and $s_{k}\in\mathbb{C}$ is the data symbol for MU $k$ with unit power, i.e. $E[|s_{k}|^{2}]=1$. Specifically, we assume that there is a maximum transmit power constraint for RRH $l$ which is given by:
\begin{equation}
\begin{aligned}
\sum\limits_{k=1}^{K}||\mathbf{w}_{l,k}||_{2}^{2}\leq P_{l}^{MAX}, \forall l\in\mathcal{L}
\end{aligned}
\end{equation}

Moreover, we consider a quasi-static fading environment and denote the channel vector from RRH $l$ to MU $k$ as $\mathbf{h}_{l,k}\in\mathbb{C}^{N_{l}\times1}, \forall l\in \mathcal{L}, \forall k\in \mathcal{K}$. Then, the received baseband signal at MU $k$ is given by
\begin{equation} \label{resig}
\begin{aligned}
y_{k}=\underbrace{\sum\limits_{l=1}^{L}\mathbf{h}_{l,k}^{H}\mathbf{w}_{l,k}s_{k}}_{\text{Desired Signal}}+\underbrace{\sum\limits_{i\in\mathcal{K},i\neq k}\sum\limits_{l=1}^{L}\mathbf{h}_{l,k}^{H}\mathbf{w}_{l,i}s_{i}}_{\text{Interference}}+z_{k}, \forall k\in\mathcal{K}
\end{aligned}
\end{equation}
where $z_{k}\sim \mathcal{CN}(0,\sigma_{k}^{2})$ is the additive Gaussian noise. Treating the interference in Eqs.(\ref{resig}) as noise, the received signal-to-interference-plus-noise ratio (SINR) at MU $k$ is given by
\begin{equation}
\begin{aligned}
SINR_{k}=\frac{|\sum\limits_{l=1}^{L}\mathbf{h}_{l,k}^{H}\mathbf{w}_{l,k}|^{2}}{\sum\limits_{i\in\mathcal{K},i\neq k}|\sum\limits_{l=1}^{L}\mathbf{h}_{l,k}^{H}\mathbf{w}_{l,i}|^{2}+\sigma_{k}^{2}}, \forall k\in\mathcal{K}
\end{aligned}
\end{equation}
\subsection{Power Model}
We assume that each RRH in C-RAN may be in one of the two states, which are sleep (SLP) and transmitting (TRA).
Then, we define the binary variable tuple $\{a_{l},b_{l,k},\forall l\in\mathcal{L},\forall k\in\mathcal{K}\}$ as the network state, where $a_{l}\in\{0,1\}$ and $b_{l,k}\in\{0,1\}$. $a_{l}$ denotes RRH power indicator: $a_{l}=1$ indicates that the RRH $l$ is in transmitting state, and $a_{l}=0$ otherwise, $\forall l\in L$. Moreover, $b_{l,k}$ denotes user association indicator: $b_{l,k}=1$ means that the MU $k$ is served by the RRH $l$, and $b_{l,k}=0$ otherwise, $\forall k\in\mathcal{K}, \forall l\in\mathcal{L}$.
Clearly, according to the relationships among $a_{l}$, $b_{l,k}$ and $\mathbf{w}_{l,k}$, we define the user association constraints as follows:
\begin{equation} \label{rela}
\begin{cases}
\{b_{l,k}=0, \forall k\in\mathcal{K}\}\Leftrightarrow a_{l}=0,\quad\quad\forall l\in\mathcal{L}\\
b_{l,k}=0\Leftrightarrow \mathbf{w}_{l,k}=\mathbf{0},\quad\quad\forall l\in\mathcal{L},\forall k\in\mathcal{K}
\end{cases}
\end{equation}
The equivalent form of Eqs.(\ref{rela}) can be written as:
\begin{equation}
\begin{cases}
b_{l,k}\neq0, \exists k\in\mathcal{K}\Leftrightarrow a_{l}=1,\quad\quad\forall l\in\mathcal{L}\\
b_{l,k}=1\Leftrightarrow \mathbf{w}_{l,k}\neq\mathbf{0},\quad\quad\forall l\in\mathcal{L},\forall k\in\mathcal{K}
\end{cases}
\end{equation}

According to \cite{power1}, the power consumption of RRH $l$ in transmitting, denoted by $P_{l}^{TRA}$, consists of both circuit and transmit power consumption. Considering the Eqs.(\ref{rela}), $P_{l}^{TRA}$ can be expressed as:
\begin{equation} \label{p1}
\begin{aligned}
P_{l}^{TRA}&=P_{l}^{CIR}+\frac{1}{\eta_{l}}\sum\limits_{k=1}^{K}b_{l,k}\|\mathbf{w}_{l,k}\|_{2}^{2}
\\&=P_{l}^{CIR}+\frac{1}{\eta_{l}}\sum\limits_{k=1}^{K}\|\mathbf{w}_{l,k}\|_{2}^{2}
\end{aligned}
\end{equation}
where $P_{l}^{CIR}$ and $\eta_{l}$ denote the circuit power consumption and the efficiency of the radio frequency power amplifier, respectively. In addition, let the constant $P_{l}^{SLP}$ denote the power consumption when the RRH $l$ is in sleep. For Pico base station, the typical values are $P_{l}^{CIR}=6.8W$, $P_{l}^{SLP}=4.3W$, and $\eta_{l}=0.25$ \cite{power1}.

With the Eqs.(\ref{rela}) and Eqs.(\ref{p1}), the power consumption of RRH $l$, denoted by $P_{l}$, can be expressed as
\begin{equation}
\begin{aligned}
P_{l}&=a_{l}P_{l}^{TRA}+(1-a_{l})P_{l}^{SLP}
\\&=a_{l}P_{l}^{CIR}+\frac{a_{l}}{\eta_{l}}\sum\limits_{k=1}^{K}b_{l,k}\|\mathbf{w}_{l,k}\|_{2}^{2}+(1-a_{l})P_{l}^{SLP}
\\&=a_{l}P_{l}^{CIR}+\frac{1}{\eta_{l}}\sum\limits_{k=1}^{K}\|\mathbf{w}_{l,k}\|_{2}^{2}+(1-a_{l})P_{l}^{SLP}
\\&=P_{l}^{SLP}+a_{l}P_{l}^{CMS}+\frac{1}{\eta_{l}}\sum\limits_{k=1}^{K}\|\mathbf{w}_{l,k}\|_{2}^{2}, \forall l\in\mathcal{L}
\end{aligned}
\end{equation}
where $P_{l}^{CMS}=P_{l}^{CIR}-P_{l}^{SLP}$. We denote $\Phi$ as the parameter tuple $\{a_{l},b_{l,k},\mathbf{w}_{l,k}, \forall l\in\mathcal{L},\forall k\in\mathcal{K}\}$. Then, by omitting the constant term $P_{l}^{SLP}$, we define the network power consumption function $F(\Phi)$ as
\begin{equation} \label{objective}
\begin{aligned}
F(\Phi)=\sum\limits_{l=1}^{L}(a_{l}P_{l}^{CMS}+\frac{1}{\eta_{l}}\sum\limits_{k=1}^{K}\|\mathbf{w}_{l,k}\|_{2}^{2})
\end{aligned}
\end{equation}
\section{Problem Formulation and Transformation}
Based on the power consumption model, we formulate here the network power minimization problem in C-RANs as a MINLP, which jointly consider the QoS requirements of MUs, maximum transmit power and the fronthaul limitation of each RRH.
\subsection{Problem Formulation}
Similar to \cite{QoS1,QoS2,QoS3,QoS4}, we employ the QoS constraints for MUs as follows:
\begin{equation}
\begin{aligned}
SINR_{k}\geq\gamma_{k}, \forall k\in\mathcal{K},
\end{aligned}
\end{equation}
where $\gamma_{k}>0$ denotes the target SINR of MU $k$.

To quantify the fronthaul cost, one obvious metric is the average bits/sec. However, considering that the information exchanged in the fronthaul includes not only user data but also signaling overhead, this metric reveals too much detail, and the associated problem is highly combinatorial \cite{Fmetric2}. Hence, similar to \cite{Introd15} and \cite{Fmetric1}, we adopt the number of active connection links which gives a first order measurement of fronthaul load as the metric of fronthaul cost instead. Then, we have the following constraints:
\begin{equation} \label{SINR}
\begin{aligned}
\sum\limits_{k=1}^{K}b_{l,k}\leq C_{l}, \forall l\in\mathcal{L}
\end{aligned}
\end{equation}
where $C_{l}$ denotes the maximum number of active connection links that RRH $l$ can serve.

With Eqs.(\ref{rela}), the QoS constraints and the fronthaul capacity constraints, the network power consumption minimization problem can be formulated as follows:
\begin{subequations}
\begin{align}
\mathscr{P}^{(pri)}&:\min F(\Phi) \\\label{1}
s.t.\quad & SINR_{k}\geq\gamma_{k}, \forall k\in\mathcal{K}\\\label{pri2}
& \sum\limits_{k=1}^{K}||\mathbf{w}_{l,k}||_{2}^{2}\leq P_{l}^{MAX}, \forall l\in\mathcal{L} \\\label{pri3}
& \sum\limits_{k=1}^{K}b_{l,k}\leq C_{l}, \forall l\in\mathcal{L} \\ \label{2}
& a_{l}=0\Leftrightarrow \{b_{l,k}=0,\forall k\in\mathcal{K}\},\forall l\in\mathcal{L} \\ \label{3}
& b_{l,k}=0\Leftrightarrow \mathbf{w}_{l,k}=\mathbf{0},\forall l\in\mathcal{L},\forall k\in\mathcal{K} \\ \label{4}
& a_{l}\in\{0,1\},b_{l,k}\in\{0,1\},\forall l\in\mathcal{L},\forall k\in\mathcal{K}
\end{align}
\end{subequations}
For the rest of this paper, we assume that problem $\mathscr{P}^{(pri)}$ is always feasible when all RRHs are active, i.e. the QoS requirement of each MU will be satisfied if $a_{l}=1, \forall l\in \mathcal{L}$.

In the following subsection, we will transform problem $\mathscr{P}^{(pri)}$ into a Mixed-Integer Second-Order Cone Program(MI-SOCP)\cite{MISOCP1}, since the problem with the user association constraints ( i.e. Eqs.(\ref{2}) and Eqs.(\ref{3})) is hard to be solved.
\subsection{Problem Transformation}
We first handle Eqs. (\ref{1}). Since the phases of $\mathbf{w}_{l,k}$ do not affect the objective function and Eqs. (\ref{1}), the QoS constraints could be rewritten as the following second order cone (SOC) constraints\cite{QoS1}:
\begin{subequations}
\begin{align}
\label{5} &\sqrt{\sum\limits_{i\neq k}|\sum\limits_{l=1}^{L}\mathbf{h}_{l,k}^{H}\mathbf{w}_{l,i}|^{2}+\sigma_{k}^{2}}\leq
\frac{1}{\sqrt{\gamma_{k}}}\mathrm{Re}\{\sum\limits_{l=1}^{L}\mathbf{h}_{l,k}^{H}\mathbf{w}_{l,k}\}, \forall k\in\mathcal{K} \\
\label{6} &\qquad\qquad\qquad\mathrm{Im}\{\sum\limits_{l=1}^{L}\mathbf{h}_{l,k}^{H}\mathbf{w}_{l,k}\}=0, \forall k\in\mathcal{K}
\end{align}
\end{subequations}

Then, we define problem $\mathscr{P}^{(pri)}$ which is a MI-SOCP:
\begin{subequations}
\begin{align}
\mathscr{P}^{(ref)}:&\min \hat{F}(\Phi)=F(\Phi)+\frac{\zeta}{L \cdot K}\sum\limits_{l=1}^{L}\sum\limits_{k=1}^{K}b_{l,k} \\
s.t.\quad&(\ref{5}),(\ref{6})\\ \label{ref1}
& \sum\limits_{k=1}^{K}||\mathbf{w}_{l,k}||_{2}^{2}\leq P_{l}^{MAX}, \forall l\in\mathcal{L} \\\label{ref3}
& \sum\limits_{k=1}^{K}b_{l,k}\leq a_{l}C_{l}, \forall l\in\mathcal{L} \\ \label{ref2}
& ||\mathbf{w}_{l,k}||_{2}\leq b_{l,k}\sqrt{P_{l}^{MAX}},\forall l\in\mathcal{L},\forall k\in\mathcal{K} \\ \label{integer}
& a_{l}\in\{0,1\},b_{l,k}\in\{0,1\},\forall l\in\mathcal{L},\forall k\in\mathcal{K}
\end{align}
\end{subequations}
where $\zeta$ is a constant with $\zeta\rightarrow 0$.

Let the parameter tuples $\Phi^{(pri)}=\{a_{l}^{(pri)},b_{l,k}^{(pri)},\mathbf{w}_{l,k}^{(pri)},\linebreak\forall l\in\mathcal{L},\forall k\in\mathcal{K}\}$ and $\Phi^{(ref)}=\{a_{l}^{(ref)},b_{l,k}^{(ref)},\mathbf{w}_{l,k}^{(ref)},\forall l\in\mathcal{L},\linebreak\forall k\in\mathcal{K}\}$
denote optimal solutions of problem $\mathscr{P}^{(pri)}$ and problem $\mathscr{P}^{(ref)}$, respectively. In what follows, we will reveals the relationship between problem $\mathscr{P}^{(pri)}$ and problem $\mathscr{P}^{(ref)}$.

$\mathbf{Lemma 1:}$ $\Phi^{(ref)}$ is a feasible solution of problem $\mathscr{P}^{(pri)}$. In addition, $\Phi^{(pri)}$ is feasible in problem $\mathscr{P}^{(ref)}$.

$\mathbf{Proof:}$ Please refer to the Appendix A.

$\mathbf{Theorem 1:}$ $\Phi^{(ref)}$ is a good approximate solution of problem $\mathscr{P}^{(pri)}$. More specifically, we have
\begin{equation}
\begin{aligned}
0\leq F(\Phi^{(ref)})-F(\Phi^{(pri)})\leq\zeta.
\end{aligned}
\end{equation}

$\mathbf{Proof:}$ Please refer to the Appendix B.

\section{The Low Complexity Inflation Algorithm}
In recent years, some commercial software packages such as CPLEX\cite{CPLEX} and MOSEK\cite{MOSEK} have been adopted to determine the optimal solution of MI-SOCP via the branch-and-cut method\cite{MISOCP2}. However, the computational complexity may be prohibitive for densely deployed scenarios in practice. Therefore, in this section, we adopt a low complexity inflation algorithm introduced by \cite{Algo1} to obtain the suboptimal solution of problem $\mathscr{P}^{(ref)}$ based on the continuous relaxation of this problem.

By relaxing the integer constraints such as Eqs.(\ref{integer}), we could transform a MI-SOCP into a SOCP, which provides a local lower bound on the optimal objective value of the original problem\cite{MISOCP1}. Then, the continuous relaxation of problem $\mathscr{P}^{(ref)}$ can be expressed as:
\begin{subequations}
\begin{align}
\mathscr{P}^{(con)}:&\min \hat{F}(\Phi) \\
s.t.\quad&(\ref{5}),(\ref{6}),(\ref{ref1})-(\ref{ref2})\\
& a_{l}\in [0,1] ,b_{l,k}\in [0,1],\forall l\in\mathcal{L},\forall k\in\mathcal{K}
\end{align}
\end{subequations}
We denote $\Phi^{(con)}=\{a_{l}^{(con)},b_{l,k}^{(con)},\mathbf{w}_{l,k}^{(con)},\forall l\in\mathcal{L},\forall k\in\mathcal{K}\}$ as the optimal solution of problem $\mathscr{P}^{(con)}$.

In the following, we will describe the low complexity inflation algorithm. Let the parameter tuple $\{a_{l}^{(n)},b_{l,k}^{(n)},\forall l\in\mathcal{L},\forall k\in\mathcal{K}\}$ and $\hat{F}^{(n)}$ denote the network state and optimal objective value in the $n$th iteration, respectively. The initialization of the algorithm is configured as $a_{l}^{(0)}=0, b_{l,k}^{(0)}=0, \forall l\in\mathcal{L},\forall k\in\mathcal{K}$, and $\hat{F}^{(0)}=\sum\limits_{l=1}^{L}(P_{l}^{CMS}+\frac{1}{\eta_{l}}P_{l}^{MAX})+\zeta$, which means that the network is shut down at the beginning of iteration and there is no $\hat{F}^{(n)},n=1,2,\ldots$ cloud be larger than $\hat{F}^{(0)}$. In addition, $\mathcal{U}^{(n)}$ denotes the indeterminate RRH-MU pair set with $\mathcal{U}^{(0)}=\{(l,k)|\forall l\in\mathcal{L},\forall k\in\mathcal{K}\}$. We gradually set one of zero-valued variables in $\{b_{l,k}^{(n-1)},\forall (l,k)\in\mathcal{U}^{(n-1)}\}$ to one in the $n$th iteration.

It is obvious that how to select the proper RRH-MU pair in the set $\mathcal{U}^{(n-1)}$
is critical for the performance of the inflation algorithm. Jointly considering the contribution to the network performance and the fronthaul limitation of each RRH, the priority level of each pair, denoted as $\alpha_{l,k}$, can be defined as:
\begin{equation} \label{priority}
\begin{aligned}
\alpha_{l,k}=\frac{|\mathbf{h}_{l,k}^{H}\mathbf{w}_{l,k}^{(con)}|^{2}}{\sum\limits_{i\neq k}|\mathbf{h}_{l,i}^{H}\mathbf{w}_{l,k}^{(con)}|^{2}}\cdot\frac{C_{l}}{\sum\limits_{j=1}^{L}C_{j}}, \forall (l,k)\in\mathcal{U}^{(0)}
\end{aligned}
\end{equation}
where $|\mathbf{h}_{l,k}^{H}\mathbf{w}_{l,k}^{(con)}|^{2}$ represents the useful signal power, and $\sum\limits_{i\neq k}|\mathbf{h}_{l,i}^{H}\mathbf{w}_{l,k}^{(con)}|^{2}$ is the interference to other MUs. According to the previous assumption, problem $\mathscr{P}^{(con)}$ is feasible with various QoS requirements, therefore, the priority level of each pair can be always obtained by calculating Eq.(\ref{priority}). In $n$th iteration, the $(l^{*},k^{*})$ denotes the RRH-MU pair with largest priority level in $\mathcal{U}^{(n-1)}$, and we will set $b_{l^{*},j^{*}}^{(n)}=1$ and $a_{l^{*}}^{(n)}=1$. Then, we will remove $(l^{*},j^{*})$ from $\mathcal{U}^{(n-1)}$. Moreover, if the number of MUs served by RRH $l^{*}$ reaches the capacity, all the RRH-MU pairs except $(l^{*},k^{*})$ will be removed from  $\mathcal{U}^{(n-1)}$.

When parameter tuple $\{a_{l}^{(n)},b_{l,k}^{(n)},\forall k\in\mathcal{K},\forall l\in\mathcal{L}\}$ in the $n$th iteration is fixed, the MI-SOCP $\mathscr{P}^{(ref)}$ is transformed into a SOCP, which can be expressed as:
\begin{subequations}
\begin{align}
\mathscr{P}^{(n)}:&\min \hat{F}(\Phi) \\
s.t.\quad&(\ref{5}),(\ref{6})\\
& \sum\limits_{k=1}^{K}||\mathbf{w}_{l,k}||_{2}^{2}\leq a_{l}^{n}P_{l}^{MAX}, \forall l\in\mathcal{L}\\
& \mathbf{w}_{l,k}=b_{l,k}^{n}\mathbf{w}_{l,k},\forall l\in\mathcal{L},\forall k\in\mathcal{K}
\end{align}
\end{subequations}
If problem $\mathscr{P}^{(n)}$ is infeasible, we will set $\hat{F}^{(n)}=\hat{F}^{(0)}$ and proceed to the next iteration. If problem $\mathscr{P}^{(n)}$ is feasible and $\hat{F}^{(n)}>\hat{F}^{(n-1)}$, we will stop and go back to the previous iteration, i.e., we set $\hat{F}^{(n)}=\hat{F}^{(n-1)}$, $b_{l^{*},k^{*}}^{(n)}=0$ and $a_{l^{*}}^{(n)}=\max\{b_{l^{*},k},\forall k\in\mathcal{K}\}$. Otherwise, we proceed to the next iteration. The pseudo-code of the low complexity inflation algorithm is presented in Algorithm $1$.
\begin{algorithm}
\caption{Low Complexity Inflation Algorithm}
  \label{alg1}
  \begin{algorithmic}[1]
  \STATE \textbf{Initialization:} $\mathcal{U}^{(0)}=\{(l,k)|,\forall l\in\mathcal{L},\forall k\in\mathcal{K}\}$; $a_{l}^{(0)}=0, b_{l,k}^{(0)}=0,\alpha_{l,k}\forall (l,k)\in\mathcal{U}^{(0)}$; a sufficiently large $\hat{F}^{(0)}$; $n=1$.
\WHILE{$\mathcal{U}^{n-1}$ is non-empty set}
\STATE{Compute $(l^{*},k^{*})= \operatorname*{argmax}\limits_{(l,k)\in\mathcal{U}^{(n-1)}}\alpha_{l,k}$.}
\STATE{Set $b_{l,k}^{(n)}=b_{l,k}^{(n-1)},a_{l}^{(n)}=a_{l}^{n-1}$.}
\STATE{Set $b_{l^{*},k^{*}}^{(n)}=1,a_{l^{*}}^{(n)}=1$.}
\STATE{Update $\mathcal{U}^{(n)}=\mathcal{U}^{(n-1)}\setminus\{(l^{*},k^{*})\}$.}
\IF{$\sum\limits_{j=1}^{K}b_{l^{*},k}^{(n)}=C_{l^{*}}$}
\STATE{$\mathcal{U}^{(n)}=\mathcal{U}^{(n)}\setminus\{(l^{*},k)|(l^{*},k)\in\mathcal{U}^{(n)},\forall k\in\mathcal{K},k\neq k^{*}\}$.}
\ENDIF
\IF{Problem $\mathscr{P}^{(n)}$ is infeasible.}
\STATE{Set $\hat{F}^{(n)}=\hat{F}^{(0)}$.}
\ELSE
\IF{$\hat{F}^{(n)}>\hat{F}^{(n-1)}$}
\STATE{Set $\hat{F}^{(n)}=\hat{F}^{(n-1)}$.}
\STATE{Set $b_{l^{*},k^{*}}^{(n)}=0, a_{l^{*}}^{(n)}=\max\limits_{k\in\mathcal{K}} b_{l^{*},k}$.}
\ENDIF
\ENDIF
\STATE{The iteration number $n=n+1$.}
\ENDWHILE
  \end{algorithmic}
\end{algorithm}

The computational complexity of the inflation algorithm in Algorithm $1$ mainly consists in solving $(\sum\limits_{l\in\mathcal{L}}C_{l}+1)$ SOCP problems, since a RRH can not serve any MU if the fronthaul reaches its limitation. For each of the SOCP problem $\mathscr{P}^{(n)}$, the computational complexity is $\mathcal{O}((K\sum\limits_{l\in\mathcal{L}}N_{l})^{3.5})$ by using the interior-point method\cite{Boyd}. Therefore, Algorithm $1$ is a polynomial time algorithm and it converges in finite iterations.
\section{Simulation Results and Discussions}
\begin{table}[!hbp]
\renewcommand\arraystretch{1.00}
\setlength{\abovecaptionskip}{5pt}
\setlength{\belowcaptionskip}{10pt}
\centering
\caption {Simulation Parameters}\label {tab:ND}
\begin{tabular}{c|c}
\toprule
\tabincell{c}{Parameter}&\tabincell{c}{Value}\\
\midrule
\hline
Path-loss at distance $d$ (km)&$148.1+37.6log_{10}(d)$ dB\\
\hline
Standard deviation of log-norm shadowing&$8$ dB\\
\hline
Small-scale fading distribution&$\mathcal{CN}(\mathbf{0,I})$\\
\hline
Noise power density&$-174$ dBm/Hz\\
\hline
System bandwidth&$10$MHz\\
\hline
Maximum transmit power of RRH $P_{l}^{MAX}$&$10$W\\
\hline
Transmitting antenna power gain&$9$ dBi\\
\hline
Constant $\zeta$&$10^{-3}$\\
\hline
\bottomrule
\end{tabular}
\end{table}
In this section, we simulate the performance of Algorithm $1$. For comparison, the scheme adopted by LTE-A is used as a benchmark, where the network is divided into non-overlapping clusters according to the locations of RRHs and MUs, and the RRHs in each cluster coordinately serve all the users within the coverage area \cite{Introd16}. We consider a network comprising $L=10$ $2$-antenna RRHs and $K=10$ or $K=15$ single-antenna MUs uniformly distributed in the square region $[-1500\;1500]\times[-1500\;1500]$ meters. The channel model is shown in Table \ref{tab:ND}. The numerical results are averaged over $100$ randomly generated network realization. According to \cite{power1}, we set $P_{l}^{CIR}=6.8W$, $P_{l}^{SLP}=4.3W$ and $\eta_{l}=0.25$.

\begin{figure}[!t]
\centering
\includegraphics[width=3.85in]{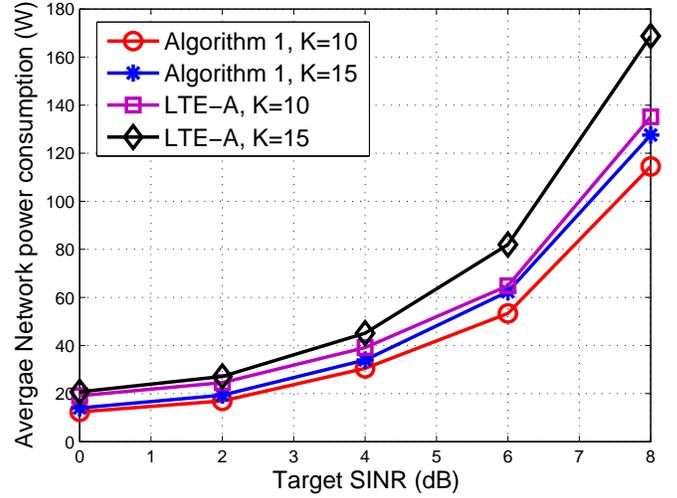}
\caption{Network power consumption versus Target SINR.}\label {tab:pic1}
\end{figure}
\begin{figure}[!t]
\centering
\includegraphics[width=3.85in]{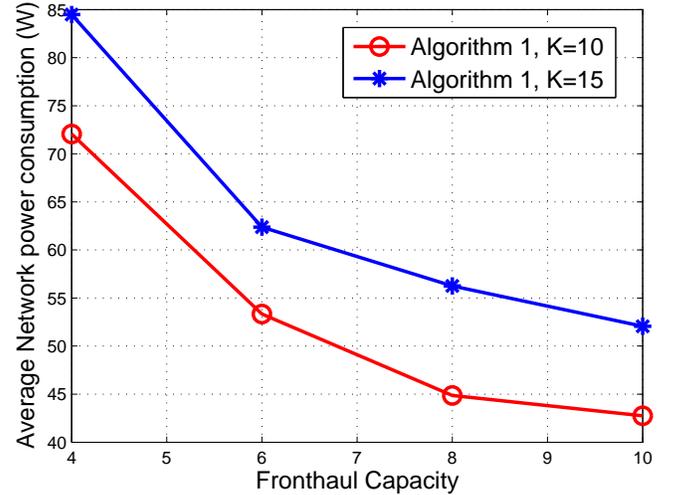}
\caption{Network power consumption versus Fronthaul capacity.}\label {tab:pic3}
\end{figure}
\begin{figure}[!t]
\centering
\includegraphics[width=3.85in]{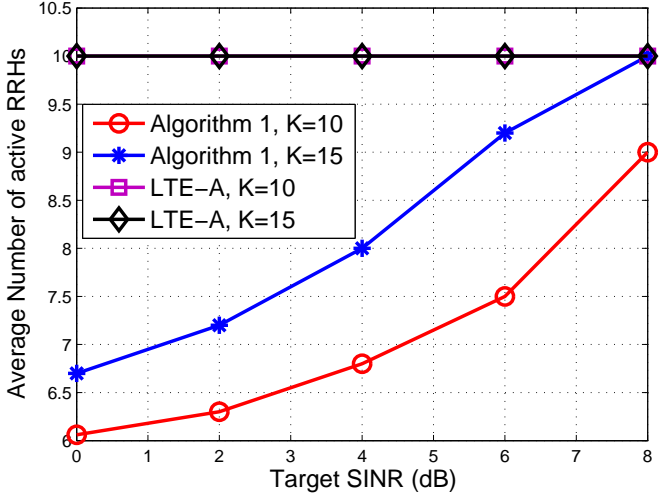}
\caption{Number of active RRHs versus Target SINR.}\label {tab:pic2}
\end{figure}
\begin{figure}[!t]
\centering
\includegraphics[width=3.85in]{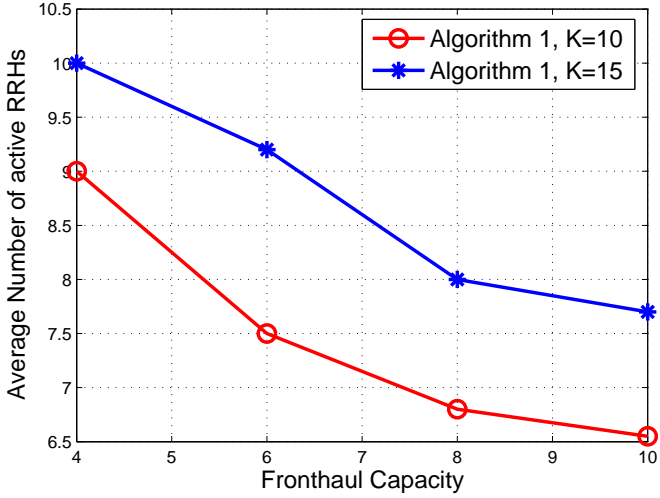}
\caption{Number of active RRHs versus Fronthaul capacity.}\label {tab:pic4}
\end{figure}

Fig.\ref{tab:pic1} shows the variation in network power consumption versus target SINR when fronthaul capacity equals $6$. We observe from this figure that the performance of Algorithm $1$ is better than that of LTE-A beamforming algorithm and the gap between the two algorithms increases with the target SINR, since the RRHs serving each MU in Algorithm $1$ are overlapping and adopt CoMP to jointly transmit data, which converts the ICI into useful signals. In addition, it is easy to understand that the network consumption with $K=10$ is lower than that with $K=15$, since higher transmitpower and more active RRHs are required to serve the excess MUs.

Fig.\ref{tab:pic3} depicts the variation in network power consumption versus fronthaul capacity when target SINR is $6dB$. As we can see, the network power consumption decreases with the fronthaul capacity. This is because the RRH can serve more MUS and more diversity gain can be obtained. Moreover, the curves descends slowly when each RRH can serve more MUs, as the fronthaul capacity is not the primary limitation on the network performance.

Fig.\ref{tab:pic2} and Fig.\ref{tab:pic4} show the variation in number of active RRHs versus target SINR and fronthaul capacity, respectively. The fronthaul capacity and target SINR are $6$ and $6dB$ in the two figures, respectively. These two figures illustrate that more RRHs have to be active with high SINR requirement or small fronthaul capacity, and more RRHs can be in sleep if there are fewer MUs when we adopt Algorithm $1$. In addition, Fig.\ref{tab:pic2} shows that all RRHs are active in LTE-A beamforming algorithm, even if the number of MUs is small and the target SINR is low.

\section{Conclusion}
In this paper, we proposed a joint user association and downlink beamforming scheme for C-RANs with limited fronthaul to minimize the network power consumption. To be more specific, the design problem is formulated as a mixed integer nonlinear program at first. We then transform the problem into a MI-SOCP which is a SOCP when the integer variables are fixed. At last, we adopt a low complexity inflation algorithm to obtain the suboptimal solution. According to the simulation results, it has been observed that the adopted algorithm can effectively lower the network consumption compared with the scheme used by LTE-A. Since too many users served by the network will incurs problem $\mathscr{P}^{(pri)}$ is infeasible, our future work is in progress to consider the user admission control in the proposed scheme.
\section*{APPENDIX A}
\section*{Proof OF Lemma 1}
We can easily observe that the feasible region of problem $\mathscr{P}^{(pri)}$ is a subset of that of problem $\mathscr{P}^{(ref)}$. Hence, $\Phi^{(pri)}$ is a feasible solution of problem $\mathscr{P}^{(ref)}$. In addition, $\Phi^{(ref)}$ satisfies the constraints (\ref{1})-(\ref{pri3}) and (\ref{4}). Accordingly, in the following, we will prove $\Phi^{(ref)}$ satisfies the constraints (\ref{2}) and (\ref{3}).

In case that $a_{l}^{(ref)}=0$, the equalities $\{b_{l,k}^{(ref)}=0,\forall k\in\mathcal{K}\}$ hold according to constraints (\ref{ref3}). In turn, if $\{b_{l,k}^{(ref)}=0,\forall k\in\mathcal{K}\}$, we can easily know that both $a_{l}=0$ and $a_{l}=1$  are feasible. If $a_{l}^{(ref)}=1$, we have $\hat{F}(\Phi^{(ref)})|_{a_{l}^{(ref)}=1}>\hat{F}(\Phi^{(ref)})|_{a_{l}=0}$, which is in contradiction to the assumption that $\Phi^{(ref)}$ is the optimal solution of problem $\mathscr{P}^{(ref)}$. Hence, we have $a_{l}^{(ref)}=0$. Moreover, through Eqs.(\ref{ref2}), we have $\mathbf{w}_{l,k}^{(ref)}=\mathbf{0}$ if $b_{l,k}^{(ref)}=0$. We adopt similar contradicting argument to prove that $b_{l,k}^{(ref)}$ equals $0$ if $\mathbf{w}_{l,k}^{(ref)}=\mathbf{0}$ as well. Therefore, $\Phi^{(ref)}$ is feasible in problem $\mathscr{P}^{(pri)}$.

The proof is completed.
\section*{APPENDIX B}
\section*{Proof OF Theorem 1}
With the $\mathbf{Lemma 1}$, we have
\begin{equation}
\begin{aligned}
&\hat{F}(\Phi^{(ref)})-\hat{F}(\Phi^{(pri)})\leq 0 \\
&(F(\Phi^{(ref)})-F(\Phi^{(pri)}))+\frac{\zeta}{L \cdot K}\sum\limits_{l=1}^{L}\sum\limits_{k=1}^{K}(b_{l,k}^{(ref)}-b_{l,k}^{(pri)})\leq 0 \\
&F(\Phi^{(ref)})-F(\Phi^{(pri)})\leq\frac{\zeta}{L \cdot K}\sum\limits_{l=1}^{L}\sum\limits_{k=1}^{K}(b_{l,k}^{(pri)}-b_{l,k}^{(ref)})
\end{aligned}
\end{equation}

$\Phi^{(pri)}$ is a optimal solution of problem $\mathscr{P}^{(pri)}$, so
\begin{equation}
\begin{aligned}
&0\leq F(\Phi^{(ref)})-F(\Phi^{(pri)})\leq\frac{\zeta}{L \cdot K}\sum\limits_{l=1}^{L}\sum\limits_{k=1}^{K}(b_{l,k}^{(pri)}-b_{l,k}^{(ref)})\\
&\leq \frac{\zeta}{L \cdot K}\cdot L\cdot K = \zeta
\end{aligned}
\end{equation}

The proof is completed.
%

\ifCLASSOPTIONcaptionsoff
  \newpage
\fi

\bibliography{Ref}

\begin{thebibliography}{10}
\providecommand{\url}[1]{#1}
\csname url@samestyle\endcsname
\providecommand{\newblock}{\relax}
\providecommand{\bibinfo}[2]{#2}
\providecommand{\BIBentrySTDinterwordspacing}{\spaceskip=0pt\relax}
\providecommand{\BIBentryALTinterwordstretchfactor}{4}
\providecommand{\BIBentryALTinterwordspacing}{\spaceskip=\fontdimen2\font plus
\BIBentryALTinterwordstretchfactor\fontdimen3\font minus
  \fontdimen4\font\relax}
\providecommand{\BIBforeignlanguage}[2]{{%
\expandafter\ifx\csname l@#1\endcsname\relax
\typeout{** WARNING: IEEEtran.bst: No hyphenation pattern has been}%
\typeout{** loaded for the language `#1'. Using the pattern for}%
\typeout{** the default language instead.}%
\else
\language=\csname l@#1\endcsname
\fi
#2}}
\providecommand{\BIBdecl}{\relax}
\BIBdecl

\bibitem{Introd4}
A.~Checko, H.~L. Christiansen, Y.~Yan, and et~al., ``{Cloud RAN} for mobile
  networks {-} a technology overview,'' \emph{IEEE Communications Surveys
  Tutorials}, vol.~17, no.~7, pp. 405--426, 2015.

\bibitem{Introd3}
``C-ran{:} the road towards green {RAN},'' \emph{White Paper by China Mobile,
  ver. 2.5}, Oct. 2011.

\bibitem{power3}
J.~Wu, Y.~Bao, G.~Miao, S.~Zhou, and Z.~Niu, ``Base station sleeping control
  and power matching for energy-delay tradeoffs with bursty traffic,''
  \emph{IEEE Transactions on Vehicular Technology}, vol.~PP, no.~99, pp. 1--1,
  2015.

\bibitem{Introd11}
Y.~Shi, J.~Zhang, and K.~B. Letaief, ``Group sparse beamforming for green
  {Cloud-RAN},'' \emph{IEEE Transactions on Wireless Communications}, vol.~13,
  no.~5, pp. 2809--2823, May 2014.

\bibitem{nof1}
J.~Cheng, Y.~Shi, B.~Bai, W.~Chen, J.~Zhang, and K.~B. Letaief, ``Group sparse
  beamforming for multicast green cloud-ran via parallel semidefinite
  programming,'' in \emph{Communications (ICC), 2015 IEEE International
  Conference on}, June 2015, pp. 1886--1891.

\bibitem{nof2}
X.~Wang, S.~Thota, M.~Tornatore, S.-S. Lee, H.-H. Lee, S.~Park, and
  B.~Mukherjee, ``Green virtual base station in optical-access-enabled
  cloud-ran,'' in \emph{Communications (ICC), 2015 IEEE International
  Conference on}, June 2015, pp. 5002--5006.

\bibitem{front1}
M.~Peng, C.~Wang, V.~Lau, and V.~Poor, ``Fronthaul{-}constrained cloud radio
  access networks{:} insights and challenges,'' \emph{IEEE Wireless
  Communications}, vol.~22, no.~2, pp. 152--160, April 2015.

\bibitem{Introd13}
B.~Dai and W.~Yu, ``Sparse beamforming for limited{-}backhaul network mimo
  system via reweighted power minimization,'' in \emph{Proc. 2013 IEEE Global
  Communications Conference}, Dec. 2013, pp. 1962--1967.

\bibitem{Introd14}
V.~N. Ha and L.~B. Le, ``Energy{-}efficient coordinated transmission for
  {Cloud-RANs}{:} algorithm design and trade{-}off,'' in \emph{Proc. 2014 48th
  Annual Conference on Information Sciences and Systems}, Mar. 2014, pp. 1--6.

\bibitem{Introd15}
------, ``Joint coordinated beamforming and admission control for fronthaul
  constrained {Cloud-RANs},'' in \emph{Proc. 2014 IEEE Global Communications
  Conference}, Dec. 2014, pp. 4054--4059.

\bibitem{circuit}
Z.~Niu, ``Tango: traffic-aware network planning and green operation,''
  \emph{IEEE Wireless Communications}, vol.~18, no.~5, pp. 25--29, October
  2011.

\bibitem{power1}
V.~Giannini, C.~Desset, I.~Godor, P.~Skillermark, M.~Olsson, M.~Imran,
  D.~Sabella, M.~Gonzalez, O.~Blume, and A.~Fehske, ``Howmuch energy is needed
  to run a wireless network{?}'' \emph{IEEE Wireless Communications}, vol.~18,
  no.~5, pp. 40--49, Oct. 2011.

\bibitem{QoS1}
M.~Bengtsson and B.~Ottersten, \emph{Optimal and Suboptimal Transmit
  Beamforming}.\hskip 1em plus 0.5em minus 0.4em\relax USA: CRC Press, 2001.

\bibitem{QoS2}
J.~Zhao, T.~Q. Quek, and Z.~Lei, ``Coordinated multipoint transmission with
  limited backhaul data transfer,'' \emph{IEEE Transactions on Wireless
  Communications}, vol.~12, no.~6, pp. 2762--2775, June 2013.

\bibitem{QoS3}
H.~Dahrouj and W.~Yu, ``Coordinated beamforming for the multicell multi-antenna
  wireless system,'' \emph{IEEE Transactions on Wireless Communications},
  vol.~9, no.~5, pp. 1748--1759, May 2010.

\bibitem{QoS4}
S.~Luo, R.~Zhang, and T.~J. Lim, ``Downlink and uplink energy minimization
  through user association and beamforming in {C-RAN},'' \emph{IEEE
  Transactions on Wireless Communications}, vol.~14, no.~1, pp. 494--508, Jan.
  2015.

\bibitem{Fmetric2}
A.~Sanderovich, O.~Somekh, H.~V. Poor, and S.~Shamai, ``Uplink macro diversity
  of limited backhaul cellular network,'' \emph{IEEE Transactions on
  Information Theory}, vol.~55, no.~8, pp. 3457--3478, Aug 2009.

\bibitem{Fmetric1}
F.~Zhuang and V.~K.~N. Lau, ``Backhaul limited asymmetric cooperation for mimo
  cellular networks via semidefinite relaxation,'' \emph{IEEE Transactions on
  Signal Processing}, vol.~62, no.~3, pp. 684--693, Feb 2014.

\bibitem{MISOCP1}
S.~Drewes, ``Mixed integer second order cone programming,'' \emph{Ph.D.
  dissertation, Darmstadt Univ. of Technology, Darmstadt, Germany}, 2009.

\bibitem{CPLEX}
IBM, ``Using the {CPLEX} callable library,'' in \emph{Version 12}, 2009.

\bibitem{MOSEK}
\emph{Mosek ApS, www.mosek.com}, 2009.

\bibitem{MISOCP2}
P.~Bonami, M.~Kilinc, and J.~Linderoth, ``Algorithms and software for convex
  mixed integer nonlinear programs,'' \emph{in Mixed Integer Nonlinear
  Programming, ser. The IMA Vol. Math. Appl., J. Lee and S. Leyffer,Eds. New
  York, NY, USA: Springer-Verlag}, vol. 154, p. 139, 2012.

\bibitem{Algo1}
Y.~Cheng, M.~Pesavento, and A.~Philipp, ``Joint network optimization and
  downlink beamforming for {CoMP} transmissions using mixed integer conic
  programming,'' \emph{IEEE Transactions on Signal Processing}, vol.~61,
  no.~16, pp. 3972--3987, Aug. 2013.

\bibitem{Boyd}
S.~P. Boyd and L.~Vandenberghe, \emph{Convex Optimization}.\hskip 1em plus
  0.5em minus 0.4em\relax Cambridge University Press, 2004.

\bibitem{Introd16}
``Coordinated multi{-}point operation for {LTE} physical layer aspects,'' 3GPP,
  Tech. Rep. TR 36.819 V11.0.0,.

\end{thebibliography}

\end{document}